\documentclass[a4paper]{article}

\usepackage[latin1]{inputenc}
\usepackage{calc}
\usepackage{stmaryrd} 
\usepackage{amsmath,amssymb,alltt}
\usepackage{psfrag}
\usepackage{graphicx}
\usepackage{listings}
\usepackage{array}
\usepackage{xspace}
\usepackage{paralist}
\usepackage{graphicx}
\usepackage{hyperref}
\usepackage{url}

\title{The parallel implementation of the {\astree} static analyzer}
\author{David Monniaux\\
\url{David.Monniaux@ens.fr}
\url{http://www.di.ens.fr}\\
\small Centre national de la recherche scientifique (CNRS)\\
\small \mbox{École normale supérieure,}\\
\small \mbox{Laboratoire d'Informatique,}\\
\small \mbox{45 rue d'Ulm,} \mbox{75230 Paris cedex 5,} France}

\newcommand{\ccode}[1]{\textbf{#1}}    
\newcommand{\cif}{\ccode{if}}          
\newcommand{\celse}{\ccode{else}}      
\newcommand{\cwhile}{\ccode{while}}    
\newcommand{\creturn}{\ccode{return}}  
\newcommand{\cgoto}{\ccode{goto}}      
\newcommand{\ccases}{\ccode{cases}}    
\newcommand{\cbreak}{\ccode{break}}    

\newcommand{\TM}{\texttrademark}

\newcommand{\bbE}{\mathbb{E}}
\newcommand{\bbL}{\mathbb{L}}

\newcommand{\parts}[1]{\mathcal{P}(#1)}

\newcommand{\ltrue}{\mathtt{t}}
\newcommand{\lfalse}{\mathtt{f}}

\newcommand{\memdomain}{D^{\sharp}_M}
\newcommand{\memgamma}{\gamma_M}

\newcommand{\lvals}{\bbL}
\newcommand{\exprs}{\bbE}

\newcommand{\ctrlset}{L} 
\newcommand{\memset}{M}  
\newcommand{\errset}{E}  

\newcommand{\sem}[1]{\llbracket #1 \rrbracket}            
\newcommand{\semd}[1]{\llbracket #1 \rrbracket}           
\newcommand{\asemd}[1]{\llbracket #1 \rrbracket^{\sharp}} 

\newcommand{\abscommand}[1]{\textbf{#1}}
\newcommand{\aassign}{\abscommand{assign}}
\newcommand{\aguard}{\abscommand{guard}}
\newcommand{\anewvar}{\abscommand{new\_var}}
\newcommand{\adelvar}{\abscommand{del\_var}}

\newcommand{\widening}{\triangledown}

\newcommand{\astree}{\textsc{Astrée}}
\newcommand{\Astree}{{\sc Astrée}\xspace} 

\newcommand{\semantics}[1]{\ensuremath\left\llbracket\text{\texttt{#1}}\right\rrbracket}
\newcommand{\abstr}[1]{{#1}^\sharp}
\newcommand{\lfp}{\textrm{lfp}}

\begin{document}
\maketitle
\begin{abstract}
The {\astree} static analyzer is a specialized tool that can prove the
absence of runtime errors, including arithmetic overflows, in large
critical programs. Keeping analysis times reasonable for industrial
use is one of the design objectives. In this paper, we discuss the
parallel implementation of the analysis.
\end{abstract}

\section{Introduction}

The \Astree static analyzer%
\footnote{\protect\url{http://www.astree.ens.fr}}
is a tool that analyzes, fully automatically, single-threaded
programs written in a subset of the C programming language,
sufficient for many typical critical embedded programs.
The tool particularly targets control\discretionary{/}{}{/}command
applications using many floating-point variables and numerical
filters, though it has been successfully applied to other categories
of software.
It computes a super-set of the possible run-time errors.
\Astree is designed for efficiency on large software: hundreds
of thousands of lines of code are analyzed in a matter of hours,
while producing very few false alarms.
For example, some fly-by-wire avionics reactive control codes
(70\,000 and 380\,000 lines respectively, the latter of a much
more complex design) are analyzed in 1\,h and 10\,h~30' respectively
on current single-CPU PCs, with \emph{no false alarm}~
\cite{BlanchetCousotEtAl02-NJ,BlanchetCousotEtAl_PLDI03,astree:esop05}.

Other contributions
\cite{Feret:ESOP04,Mine:AST01,Mine:SAS02,Mine:ESOP04,Mine_thesis,MinePADO01}
have described the \emph{abstract domains} used in {\astree}; that is,
the data structures and algorithms implementing the symbolic
operations over abstract set of reachable states in the program to be
analyzed. However, the operations in these abstract domains must be
driven by an \emph{iterator}, which follows the control flow of
the program to be analyzed and calls the necessary operations. This
paper describes some characteristics of the iterator. We first explain
some peculiarities of our iteration algorithm as well as some
implementation techniques regarding efficient shared data
structures. These have an impact on the main contribution of the
paper, which is the parallelization technique implemented in {\astree}.

Even though {\astree} presented good enough performances to be used in
practical settings on large-scale industrial code on single-processor
systems, we designed a parallel implementation suitable both for
shared-memory multiprocessor systems and for small clusters of PCs
over local area networks. {\astree} being focused on synchronous,
statically scheduled reactive programs, we used that peculiar form of
the program to be analyzed in order to design a very simple, yet
efficient, parallelization scheme. We however show that the control-flow
properties that enable such parallel analysis are to be found in other
kinds of programs, including major classes of programs such as
event-driven user interfaces.

Section \ref{sec2:structure} describes the overall structure of the
interpreter and the most significant choices about the iteration
strategy. This defines the framework within which we implement our
parallelization scheme.
We also discuss implementation choices for some data
structures, which have a large impact on the simplicity and efficiency
of the parallel implementation.

Section \ref{sec5:parallel} describes the parallelization of the
abstract interpreter in a range of practical cases.

\section{The ASTRÉE Abstract Interpreter}\label{part:abstract_interpreter}
  \label{sec2:structure}
  Our static analyzer is structured in several hierarchical layers:
  \begin{asparaitem}
  \item a denotational-based \emph{abstract interpreter}
    abstractly executes the
    instructions in the programs by sending orders to the abstract
    domains;
  \item a {\em partitioning domain} \cite{lmxr:esop05} handles the
    partitioning of traces depending on various criteria; it also
    operates the partitioning with respect to the call stack;
  \item a \emph{branching abstract domain} handles forward branching
    constructs such as forward \texttt{goto}, \texttt{break},
    \texttt{continue}; 
  \item a \emph{structure abstract domain} resolves all accesses to
    complex data structures (arrays, pointers, records...) into may- or
    must-aliases over \emph{abstract cells}
    \cite[§6.1]{BlanchetCousotEtAl_PLDI03};
  \item various numerical domains express different kinds of
    constraints over those cells; each of these domains can query
    other domains for information, and send information to those
    domains (\emph{reduction}).
  \end{asparaitem}

\subsection{A denotational-based interpreter}  
  \label{part:interpreter}
  Contrary to some presentations or examples of abstract
  interpretation-based static analysis
  \cite{CousotCousot77} , we did not choose
  to obtain results through the direct resolution (or
  over-approximation) of a system of semantic equations, but rather to
  follow the denotational semantics of the program as in
  \cite[Sect.~13]{Cousot_Marktoberdorf}. 

  Consider the following fragment of the C~programming language:

  \( \qquad
  \begin{array}{lcl} 
    l
    & ::=
    & x \mid t \lbrack e \rbrack  \,\mid \, \ldots
    \qquad \text{l-values}
    \\
    e
    & ::=
    & l \mid e \oplus e \, \mid \, \ominus e \mid \ldots
    \qquad \text{expressions } (\oplus \in \{ +, \star, \ldots \};
    \ominus \in \{ -, \ldots \})
    \\
    s
    & ::=
    & \tau \; x; \, \mid \, l \, = \, e \, \mid \,
    \cif( e ) \{ s; \ldots; s \} \celse \{ s; \ldots; s \}
    \, \mid \, \cwhile( e ) \{ s; \ldots; s \}
    \\
  \end{array}
  \)
  
  \( \ctrlset \) is the set of control states, \( \tau \) is any type,
  \( \bbL \) (resp. \( \bbE \)) is the set of l-values (resp. expressions).
  The concrete semantics of a statement \( s \) is
  a function \( \semd{s}: \memset \rightarrow \parts{\memset} \times
  \parts{\errset} \) where \( \memset \) (resp. \( \errset \)) is
  the set of memory states (resp. errors).
  Given an abstraction of sets of stores defined by an abstract domain
  \( \memdomain \) and a concretization function \( \memgamma: \memdomain
  \rightarrow \parts{\memset} \), we can derive an approximate abstract
  semantics \( \asemd{P} : \memdomain \rightarrow \memdomain \) of
  program fragment \( P \)
  by following the methodology of abstract interpretation
  \cite{Cousot_JLP92}.

  The soundness of \( \asemd{P} \) can be stated as follows:
  if \( \semd{P}( \rho ) = (m_0,e_0) \) and \( \rho \in \memgamma(\abstr{d}) \),
  then \( m_0 \subseteq \memgamma( \asemd{P}(\abstr{d}) ) \) (resp. for the
  error list).
  The principle of the interpreter is to compute \( \asemd{P}(
  \abstr{d} ) \) by induction on the syntax of~$P$.
  \( \memdomain \) should provide abstract counterparts (\( \aassign,
  \anewvar, \adelvar, \aguard \)) to the concrete operations (assignment,
  variable creation and deletion, condition test).
  For instance, \( \aassign \) should be a function in \( \bbL
  \times \bbE \times \memdomain \rightarrow \memdomain \), that inputs
  an l-value, an expression
  and an abstract value and returns a sound over-approximation
  of the set of stores resulting from the assignment:
  \(
  \forall l \in \bbL, \allowbreak
  \forall e \in \bbE, \allowbreak
  \forall \abstr{d} \in \memdomain,\allowbreak
  \; \{ \rho \lbrack \sem{l}( \rho ) \mapsto \sem{e}( \rho ) \rbrack
  \mid \rho \in \memgamma( \abstr{d} ) \}
  \subseteq \memgamma( \aassign( l, e, \abstr{d} ) )
  \).
  Soundness conditions for the other operations (\( \aguard,
  \anewvar, \adelvar \)) are similar.
  \[
  \begin{array}{rcl}
    \asemd{ l = e; }( \abstr{d} )
    & = 
    & \aassign( l, e, \abstr{d} ) \quad \text{where } l \in \lvals, e \in \exprs
    \\
    \asemd{ \{ \tau \, x; s_0 \} }( \abstr{d} )
    & =
    & \adelvar( \tau \, x, \asemd{s_0}( \anewvar( \tau \, x, \abstr{d} ) ) )
    \\
    \asemd{ \cif( e ) \, s_0 \, \celse \, s_1; }( \abstr{d} )
    & =
    & \asemd{ s_0 }( \aguard( e, \ltrue, \abstr{d} ) ) \sqcup
    \asemd{ s_1 }( \aguard( e, \lfalse, \abstr{d} ) )
    \\
    \asemd{ \cwhile( e ) \, s_0 }( \abstr{d} )
    & = 
    & \aguard( e, \lfalse, \abstr{\lfp} \abstr{\phi} )
    \\
    &
    & \; \text{where: } \abstr{\phi}: \abstr{x} \in \memdomain \mapsto
    \abstr{d} \sqcup \asemd{s_0}( \aguard( e, \ltrue, \abstr{x} ) )
    \\
  \end{array}
  \]

  The function \( \abstr{\lfp} \) computes a post-fixpoint of any abstract
  function (i.e., approximation of the concrete least-fixpoint). While
  the actual scheme implemented is somewhat complex, it is sufficient
  to say that $\abstr{\lfp} \abstr{f}$ outputs some $\abstr{x}$ such
  that $\abstr{f}(\abstr{x}) \abstr{\sqsubseteq} \abstr{x}$ for some
  decidable ordering $\abstr{\sqsubseteq}$ such that
  \(\forall \abstr{x},\abstr{y}~
   \abstr{x} \abstr{\sqsubseteq} \abstr{y} \implies
   \gamma(\abstr{x}) \subseteq \gamma(\abstr{y})\).
  This abstract fixpoint is sought by the iterator in ``iteration
  mode'': possible warnings that could occur within the code are not
  displayed when encountered. Then, once $\abstr{L}=\abstr{\lfp} \abstr{f}$ is
  computed --- an invariant for the loop body ---, the iterator
  analyzes the loop body again and displays possible warnings. As a
  supplemental safety measure, we check again that
  $\abstr{f}(\abstr{L}) \abstr{\sqsubseteq} \abstr{L}$.%
\footnote{Let us note that the computationally costly part of the
  analysis is finding the loop invariant, rather than checking it.
  P.~Cousot suggested the following improvement over our existing
  analysis: using different implementations for finding the invariant
  and checking it (at present, the same program does both).
. For instance, the checking phase
  could be a possibly less efficient version, whose safety would be
  formally proved. However, since all abstract domains and most
  associated algorithms would have to be implemented in that ``safe''
  analyzer, the amount of work involved would be considerable and we
  have not done it at this point. Also, as discussed in
  Sect.~\ref{part:parallel_impl}, both implementations would have to
  yield identical results, which means that the ``safe'' analysis
  would have to mimic the ``unsafe'' one in detail.} 

  $\sqcup$ is an abstraction of the concrete union~$\cup$:
  \(
  \forall \abstr{d_1},\abstr{d_2},
  \gamma(\abstr{d_1}) \cup \gamma(\abstr{d_2}) \subseteq
  \gamma(\abstr{d_1} \sqcup \abstr{d_2})
  \).
  
  An abstract domain handles the call stack; currently in {\astree}, it amounts
  to partitioning states by the full calling
  context \cite{lmxr:esop05}. {\astree} does
  not handle recursive functions;%
  \footnote{More precisely, it can analyze recursive programs, but
    analysis may fail to terminate. If analysis terminates, then its
    results are sound.}
  this is not a problem with critical embedded code, since programming
  guidelines for such systems generally prohibit the use of recursive
  functions.
  Functions are analyzed as if they
  were inlined at the point of call.
  Multiple targets for function
  pointers are analyzed separately, and the results merged
  with~$\sqcup$; see §\ref{part:parallel} for an application to
  parallelization.
  
  Other forms of branches are dealt with by an extension of the
  abstract semantics.
  Explicit gotos are rarely used in C, except as forward branches to
  error handlers or for exiting multiple loops;
  however, semantically similar branching
  structures are very usual and include \( \ccases \) structures,
  \( \cbreak \) statements and \( \creturn \) statements.
  Indeed, a return statement \( \creturn \; e \) carries out two
  operations:
    first, it evaluates \( e \) and stores the value as the function
    result;
    then, it terminates the current function, i.e. branches to the end of
    the function.
  In this paper, we only consider the case of  forward-branching
  \( \cgoto \)'s; the other constructs then are straightforward extensions.
  
  We extend the syntax of statements with a goto statement \( \cgoto \;
  l \) where \( l \) is a program point (we implicitly assume that there
  is a label before each statement in the program).
  The execution of a statement \( s \) may yield either a new
  memory state or a branching to a point after \( s \).
  Therefore, we lift the definition of the semantics into a function
  \( \semd{s}: (\memset \times (\ctrlset \rightarrow \parts{\memset}))
  \rightarrow (\parts{\memset} \times (\ctrlset \rightarrow
  \parts{\memset}) \times \errset) \).
  The concrete states before and after each statement no longer
  consist solely of a set of memory states, but of a set of memory
  states for the ``direct'' control-flow as well as a set of memory
  states for each label $l$, representing all the memory states that
  have occurred in the past for which a forward branch to $l$ was
  requested.

  The concrete semantics of \( \cgoto \; l \) is defined by:
  \( \semd{\cgoto \; l ;}( I_i, \phi_i ) \allowbreak = \allowbreak (\bot,
    \phi_i\lbrack l \allowbreak \mapsto \allowbreak
    \phi_i(l) \cup I_i\rbrack) \)
  and the concrete semantics of a statement $s$ at label $l$
  is defined from the semantics without branches as:
  \( \semd{l : s;}( I_i, \phi_i ) \allowbreak = \allowbreak
     (\{ I_i \} \cup \phi_i( l ),\phi_i)\).
  The definition of the abstract semantics
  can be extended in the same way.
  We straightforwardly lift the abstract semantics of a
  statement \( s \) into a function \( \asemd{s}: \memdomain \times
  (L \rightarrow \memdomain) \rightarrow \memdomain \times (L
  \rightarrow \memdomain) \).
  

\subsection{Rationale and efficiency issues}
\label{part:efficiency}
  The choice of the denotational approach was made for two reasons~:
  \begin{compactitem}
  \item
    Iteration and widening techniques on general graph representations of
    programs are more complex. Essentially, these techniques partly have to
    reconstruct the natural control flow of the program so as to
    obtain an efficient propagation
    flow \cite{Bourdoncle93,HorwitzEtAl_ActaInformatica87}.
    Since our programs are
    block-structured and do not contain backward \texttt{goto}'s,
    this flow information is already present in their syntax; there is
    no need to reconstruct it. \cite[Sect.~13]{Cousot_Marktoberdorf}

  \item
    It minimizes the amount of
    memory used for storing the abstract environments.
    While our storage methods maximize the sharing between
    abstract environments, our experiments showed that storing an abstract
    environment for each program point (or even each branching point) in
    main memory was too costly. Good forward/backward
    iteration techniques do not need to store environments at that
    many points, but this measurement still was an indication that
    there would be difficulties in implementing such schemes.
  \end{compactitem}

  We measured the memory required for
  storing the local invariants at part or all of the program points, for
  three industrial control programs representative of those we are interested
  in; see the table below. We performed several measurements, depending on
  whether invariant data was saved at all statements,
  at the beginning and end of each block, and at the beginning and end
  of each function.

  For each program and measurement, we provide two figures: from left
  to right, the peak memory observed during the
  analysis, then the size of the serialized invariant (serialization is
  performed for saving to files or for parallelization purposes,
  and preserves the sharing property of the internal representation).

  Benchmarks (see below) show that keeping local invariants at the
  boundaries of every block in main memory
  is not practical on large programs;
  even restricting to the boundaries of functions results in a
  major overhead. A database system for storing invariants on
  secondary storage could be an option, but Brat and Venet have
  reported significant difficulties and complexity with that
  approach~\cite{BratVenet_AERO05}. Furthermore, such an approach
  would complicate memory sharing, and perhaps force the use of
  solutions such as  ``hash-consing'', which we have avoided so far.

  Memory requirements are expressed in megabytes; analyses
  were run in 64-bit mode on a dual Opteron 2.2~GHz machine with 8~Gb RAM.%
  \footnote{Memory requirements are smaller on 32-bit systems.}
  On many occasions, we had to abort the computation due to large
  memory requirements causing the system to swap.

  \begin{center}\footnotesize
  \begin{tabular}{|l|r|r|r|r|r|r|}
    \hline
    & \multicolumn{2}{c|}{Program~1}
    & \multicolumn{2}{c|}{Program~2}
    & \multicolumn{2}{c|}{Program~3} \tabularnewline
    \hline
    \# of lines of C code &
      \multicolumn{2}{r|}{67,553} &
      \multicolumn{2}{r|}{232,859} &
      \multicolumn{2}{r|}{412,858} \tabularnewline
    \hline
    \# of functions &
      \multicolumn{2}{r|}{650} &
      \multicolumn{2}{r|}{1,900} &
      \multicolumn{2}{r|}{2,900} \tabularnewline
    \hline
    Save at all statements
    & \( 3300 \) & \( 688 \) 
    & \( > 8000 \) & swap 
    & \( > 8000 \) & swap 
    \tabularnewline \hline
    Save at beginning / end of blocks
    & \( 2300 \) & \( 463 \) 
    & \( > 8000 \) & swap
    & \( > 8000 \) & swap
    \tabularnewline \hline
    Save at beginning / end of functions
    & \( 690 \) & \( 87 \) 
    & \( 2480 \) & \( 264 \) 
    & \( 4800 \) & \( 428 \) 
    \tabularnewline \hline
    Save main loop invariant only
    & \( 415 \) & \( 15 \) 
    & \( 1544 \) & \( 53 \) 
    & \( 2477 \) & \( 96 \) 
    \tabularnewline \hline
    No save
    & \multicolumn{2}{r|}{\( 410 \)} 
    & \multicolumn{2}{r|}{\( 1544 \)}
    & \multicolumn{2}{r|}{\( 2440 \)} 
    \tabularnewline \hline
  \end{tabular}
  {\scriptsize Benchmarks courtesy of X.~Rival.}
  \end{center}
  
  Memorizing invariants at the head
  of loops (the least set of invariants we can keep so as to be able
  to compute widening chains) thus entails much smaller memory
  requirements than a naïve graph-based implementation; the latter
  is intractable on reasonably-priced current computers on the class
  of large programs that we are interested in. It is possible
  that more complex memorization schemes may make graph-based
  algorithms tractable, but we did not investigate such schemes
  because we had an efficient and simple working system.

  Regarding efficiency, it soon became apparent that a major factor
  was the efficiency of the $\sqcup$ operation.
  In a typical program, the number of tests will be roughly
  linear in the length of the code. In the control programs that
  {\astree} targets, the number of state variables (the values of which
  are kept across iterations) is also
  roughly linear in the length $l$ of the code. This means that if the
  $\sqcup$ operation takes linear time in the number of variables
  ---~an apparently good complexity~---, an
  iteration of the analyzer takes $\Theta(l^2)$ time, which is
  prohibitive. We therefore argue that \emph{what matters is the
  complexity of $\sqcup$ with respect to the number of updated
  variables}, which should be almost linear:
  if only $n_1$ (resp. $n_2$) variables are
  touched in the \texttt{if} branch (resp. \texttt{else} branch), then
  the overall complexity should be at most roughly
  $O(n_1+n_2)$.

  We achieve such complexity with our implementation
  using balanced trees with strong memory sharing
  and ``short-cuts'' \cite[§6.2]{BlanchetCousotEtAl02-NJ}.
  Experimentation showed that memory sharing was good with the rough
  physical equality tests that we implement, without the need for much
  more costly techniques such as hash consing. Indeed, experiments
  show that considerable sharing is kept after the abstract execution
  of program parts that modify only parts of the global state
  (see the $\Delta$-compression in §\ref{part:delta}).
  Though simple, this memory-saving technique is fragile; data sharing
  must be conserved by all modules in the program.
  This obligation had an impact on the design of the
  communications between parallel processes.

\section{Parallelization}\label{part:parallel}
  \label{sec5:parallel}
In iteration mode, we analyze tests in the following way:
\(
\asemd{ \cif( e ) \, s_0 \, \celse \, s_1; }( \abstr{d} )\allowbreak
=\allowbreak
\asemd{ s_0 }( \aguard( e, \ltrue, \abstr{d} ) ) \sqcup
\asemd{ s_1 }( \aguard( e, \lfalse, \abstr{d} ) )
\).
The analyses of \( s_0 \) and \( s_1 \) may be conducted in total
separation, in different threads or
processes, or even on different machines.
Similarly, the semantics of an indirect function call may be
approximated as:
$\abstr{\semantics{(*f)();}}(\abstr{a}) =
 \bigsqcup_{g \in \semantics{f}(\abstr{a})}
 \abstr{\semantics{$g$}}(\abstr{a})$:
$g$ ranges on all the possible code blocks to which \texttt{f} may point.

\subsection{Dispatch points}\label{part:sequencers}
In usual programs, most tests split the control flow between short
sequences of execution; the overhead of analyzing such short paths in
separate processes would therefore be considerable with
respect to the length of the analysis itself. However, there exist
wide classes of programs where a few tests (at ``dispatch points'')
divide the control flow between long executions.
\label{part:parallel_motivation}
In particular, there exist several important kinds of software
consisting in a large event loop: the system waits for an event, then
a ``dispatcher'' runs an appropriate (often fairly complex) handler
routine:
\begin{tabbing}
Initialization\\
\textbf{while true do}\\
~~\=~~\=~~\=\kill
\>wait for a request $r$\\
\>\textbf{dispatch} according to the type of $r$ \textbf{in}\\
\>\>handler for requests of the first type\\
\>\>handler for requests of the second type\\
\>\>...\\
\textbf{done}
\end{tabbing}

This program structure is quite frequent in
network services and traditional graphical user interfaces
(though nowadays often wrapped inside a
callback mechanism):
\begin{tabbing}
Initialization\\
\textbf{while true do}\\
~~\=~~\=~~\=\kill
\>wait for an event $e$\\
\>\textbf{dispatch} according to $e$ \textbf{in}\\
\>\>event handler 1\\
\>\>event handler 2\\
\>\>...\\
\textbf{done}
\end{tabbing}

\label{part:reactive_scheduler}
Many critical embedded programs are also of this form.
We analyze reactive programs that, for the most part, implement in
software a directed graph of numeric filters. Those numeric filters
are in general the discrete-time counterparts of hardware,
continuous-time components, with various sampling rates. The system is
thus made of $n$ components, each clocked with a period
$p_i \cdot p$, where $1/p$ is a master clock rate (say, 1~kHz).
It is statically scheduled as a succession of ``sequencers'' numbered
from $0$ to $N-1$ where $N$ is the least common multiple of the $p_i$.
A task of period $p_i.p$ is scheduled in all sequencers numbered
$k.p_i+c_i$. $c_i$ may often be arbitrarily chosen; judicious choices of
$c_i$ allow for static load balancing, especially with respect to
worst-case execution time (all sequencers should complete within a
time of $p$). Thus, the resulting program is of the form:
\begin{tabbing}
Initialization\\
\textbf{while true do}\\
~~\=~~\=~~\=\kill
\>wait for clock tick ($1/p~\textrm{Hz}$)\\
\>\textbf{dispatch} according to $i$ \textbf{in}\\
\>\>sequencer 0\\
\>\>sequencer 1\\
\>\>...\\
\>$i := i+1 \pmod N$\\
\textbf{done}
\end{tabbing}

Our analysis is imprecise with respect to the succession of values of
$i$; indeed, it soon approximates $i$ by the full interval
$[0,N-1]$. This is equivalent to saying that any of the sequencers is
nondeterministically chosen. Yet, due to the nature of the studied
system, this is not a hindrance to proving the absence of runtime
errors: each of the $n$ subcomponents should be individually sound,
whatever the sampling rate, and the global stability of the system
does not rely on the sampling rates of the subcomponents, within
reasonable bounds.

Our system could actually handle parallelization at any place in the
program where there exist two or more different control flows, by
splitting the flows between several processors or machines; it is
however undesirable to fork processes or launch remote analyses for
simple blocks of code.
Our current system decides the splitting point according to some simple
ad-hoc criterion, but we could use some more universal criteria. For
instance, the analyzer, during the first iteration(s), could measure
the analysis times of all branches in \texttt{if}, \texttt{switch} or
multi-aliases function calls; if a control flow choice takes place
between several blocks for which the analysis takes a long time, the
analysis could be split between several machines in the following
iterations. To be efficient, however, such a system would have to do
some relatively complex workload allocation between processors and
machines; we will thus only implement it when really necessary.

\subsection{Parallelization implementation}\label{part:delta}
\label{part:parallel_impl}

\begin{figure}[htb]
\begin{center}
\includegraphics{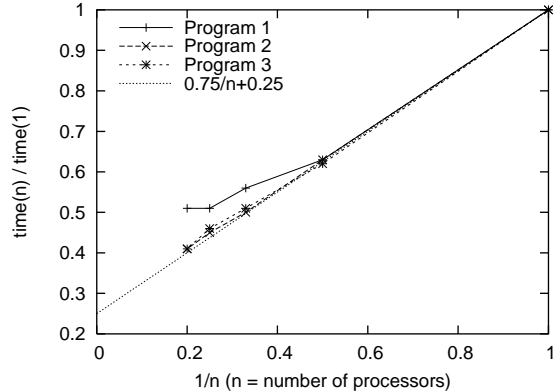}
\end{center}
\caption{Parallelization performances (dual-2.2~GHz Opteron machine +
  2~GHz AMD64 machines).}
\label{fig:parallel_benchmark}
\end{figure}

Instead of analyzing all dispatch branches in sequence, we split the
workload between $p$ several processors (possibly in different
machines).
We replace the iterative computation of
\(
\abstr{f}(\abstr{X}) =
  \abstr{\semantics{$P_1$}}(\abstr{X}) \sqcup (\abstr{\semantics{$P_2$}}(\abstr{X} \sqcup \dots
  \abstr{\semantics{$P_n$}}(\abstr{X}) ) \dots )
\)
by a parallel computation
\(
\abstr{f}(\abstr{X}) = \bigsqcup_{i=1}^p (
  \bigsqcup_{k\in\pi_i} \allowbreak \abstr{\semantics{$P_k$}}(\abstr{X}))
\) where
the $\pi_k$ are a partition of $\{1,\ldots,n\}$. Let us note $\tau_j$
the time needed to compute $\abstr{\semantics{$P_j$}}$.
$l_k=\sum_{j\in\pi_k} \tau_j$ is the time spent by processor~$i$.

For maximal efficiency, we
would prefer that the $l_i$ should be close to each other, so as to
minimize the synchronization waits. Unfortunately, the problem of
optimally partitioning into the $\pi_k$ is NP-hard even in the case
where $p=2$~\cite{Mertens04:easiest_hard_problem}. If the $\tau_i$ are
too diverse, randomly shuffling the list may yield improved
performance.
In practice, the real-time programs that we analyze are scheduled so
that all the $P_i$ have about the same worst-case execution time, so
as to ensure maximal efficiency of the embedded
processor; consequently, the $\tau_i$ are reasonably close
together and random shuffling does not bring significant improvement;
in fact, in can occasionally reduce performances.

For large programs of the class we are interested in, the analysis
times (Fig.~\ref{fig:parallel_benchmark}) for $n$ processors is
approximately $0.75/n+0.25$ times the
analysis time on one processor; thus, clusters of more than 3 or 4
processors are not much interesting:
\begin{center}
\begin{tabular}[b]{|l|r|r|r|}
\hline
& Prog~1 & Prog~2 & Prog~3 \\
\hline
\# lines & 67,553 & 232,859 & 412,858 \\
1 CPU & 26'28" & 5h~55' & 11h~30' \\
2 CPU & 16'38" & 3h~43' & 7h~09' \\
3 CPU & 14'47" & 2h~58' & 5h~50' \\
4 CPU & 13'35" & 2h~38' & 5h~06' \\
5 CPU & 13'26" & 2h~25' & 4h~44' \\
\hline
\end{tabular}
\end{center}

Venet and Brat also have experimented with parallelization
\cite[§5]{VenetBrat_PLDI04}, with similar conclusions;
however, the class of programs to be
analyzed and the expected precisions of their analysis are too
different from ours to make direct comparisons meaningful.

Because it is difficult to determine the $\tau_i$ in advance,
{\astree} features
an optional randomized scheduling strategy, which reduces
computation times on our examples by 5\%, with computation times
on 2 CPUs $\simeq$58\% of those on~1.

We reduced transmission
costs by sending only the differences between abstract values at the
input and the output --- when the remote computation is
$\abstr{\semantics{P}}(\abstr{d})$, only answer the difference between
$\abstr{d}$ and $\abstr{\semantics{P}}(\abstr{d})$. This difference is
obtained by physical comparison of data structures, excluding shared
subtrees (Sect.~\ref{part:efficiency}).
The advantage of that method is twofold:
\begin{itemize}
\item Experimentally, such ``$\Delta$-compression'' results in
transmissions of about 10\% of the full size on our
examples. This reduces transmission costs on networked implementations.
\item Recall that we make analysis tractable by sharing data
  structures (Sect.~\ref{part:efficiency}). We however enforce this
  sharing by simple pointer comparisons (i.e. we do not construct
  another copy of a node if our procedure happens to have the original
  node at its disposal), which is fast and simple but does not
  guarantee optimal sharing.
  Any data coming from the network, even though logically
  equal to some data already present in memory, will be loaded at a
  different location; thus, one should avoid merging in redundant data.
  Sending only the difference back to the master
  analyzer thus dramatically reduces the amount of unshared data
  created by networked merge operations.
\end{itemize}

We request that the $\sqcup$ operator should be associative and
commutative, so that $\abstr{f}$ does not depend on the chosen
partitioning. Such a dependency would be detrimental for two reasons:
\begin{itemize}
\item
  If the subprograms
  $P_1,\ldots,P_n$ are enclosed within a loop, the nondeterminism of
  the abstract transfer function $\abstr{f}$ complicates the analysis
  of the loop. As we said in
  Sect.~\ref{part:interpreter}, we use an ``abstract fixpoint''
  operator $\abstr{\lfp}$ that terminates when it
  finds $\abstr{L}$ such that $\abstr{f}(\abstr{L}) \sqsubseteq
  \abstr{L}$. Because this check is performed at least twice, it would be
  undesirable that the comparisons yield inconsistent results.
\item
  For debugging and end-user purposes, it is undesirable that the
  results of the analysis could vary for the same analyzer and inputs
  because of runtime vagaries.
\footnote{For the same reasons, care should be exercised in networked
  implementations so that different platforms output the same analysis
  results on the same inputs. Subtle problems may occur in that
  respect; for instance, there may be differences between
  floating-point implementations.
  We use the native floating-point
  implementation of the analysis platform; even though all our host
  platforms are IEEE-compatible, the exact same analysis code may
  yield different results on various platforms, because
  implementations are allowed to provide more precision that requested
  by the norm. For example, the IA32\TM
  (Intel Pentium\TM) platform under Linux\TM (and some other operating
  systems) computes by default internally with 80 bits of precision
  upon values that are specified to be 64-bit IEEE double precision
  values. Thus, the result of computations on that platform may depend
  on the register scheduling done by the compiler, and may also differ
  from results obtained on platforms doing all computations on 64-bit
  floating point numbers (PowerPC\TM, and even IA32\TM and AMD64\TM
  with some code generation and system settings).
  Analysis results, in all cases, would be sound, but they would
  differ between implementations, which would be undesirable for the
  sake of reproducibility and debugging, and also for parallelization,
  as explained here.
  We thus force (if possible) the
  use of double (and sometimes single) precision IEEE floating-point
  numbers in all computations within the analyzer.} 
\end{itemize}

In this case of a loop around the $P_1,\ldots,P_n$, we could have
alternatively used asynchronous iterations \cite{Cousot_RR88}. To compute
$\lfp \abstr{f}$, one can use a central repository $\abstr{X}$,
initially containing $\bot$; then, any processor $i$ computes
$\abstr{f}_i(\abstr{X})=
  \bigsqcup_{k\in\pi_i} \abstr{\semantics{$P_k$}}(\abstr{X})$
and replaces $\abstr{X}$ with
$\abstr{X} \widening \abstr{f}_i(\abstr{X})$.
If the scheduling is fair (no $\semantics{$P_k$}$ is ignored
indefinitely), such iterations converge to an
approximation of the least fixpoint of
$X \mapsto \cup_k \semantics{$P_k$}(X)$.
However, we did not implement our analyzer this way. Apart from the
added complexity, the nondeterminism of the results was undesirable.

\section{Conclusion}
We have investigated both theoretical and practical matters regarding
the computation of fixpoints and iteration strategies for static
analysis of single-threaded, block-structured programs, and proposed
methods especially suited for the analysis of large synchronous
programs: a denotational iteration scheme, maximal data sharing between
abstract invariants, and parallelization schemes. In several
occasions, we have identified possible extensions of our system.

Two major problems we have had to deal with were long computation
times, and, more strikingly, large memory requirements, both owing to
the very large size of the programs that we consider. Additionally, we
had to keep a very high precision of analysis over complex numerical
properties in order to be able to certify the absence of runtime
errors in the industrial programs considered.

We think that several of these methods will apply to other classes of
programs. Parallelization techniques, perhaps extended, should apply
to wide classes of event-driven programs; loop iteration techniques
should apply to any single-threaded programs; data sharing and
``union'' optimizations should apply to any static analyzer. We also
have identified various issues of a generic interest with respect to
widenings and narrowings.
\medskip

\noindent\textbf{Acknowledgments:} We wish to thank P.~Cousot and X.~Rival, as
well as the rest of the {\astree} team.


\bibliographystyle{plain}
\bibliography{aplas}
\end{document}